# Active metamaterials with negative static electric susceptibility

F. Castles[1,2]*, J. A. J. Fells[3], D. Isakov[1,4], S. M. Morris[3], A. Watt[1] and P. S. Grant[1]*

[1] Department of Materials, University of Oxford, Parks Road, Oxford OX1 3PH, UK.

[2] School of Electronic Engineering and Computer Science, Queen Mary University of London, Mile End Road, London E1 4FZ, UK.

[3] Department of Engineering Science, University of Oxford, Parks Road, Oxford OX1 3PJ, UK.

[4] Warwick Manufacturing Group, University of Warwick, Coventry CV4 7AL, UK.

* email: f.castles@qmul.ac.uk; patrick.grant@materials.ox.ac.uk

**Well-established textbook arguments suggest that static electric susceptibility must be positive in "all bodies"[1]. However, it has been pointed out that media that are not in thermodynamic equilibrium are not necessarily subject to this restriction; negative static electric susceptibility has been predicted theoretically in systems with inverted populations of atomic and molecular energy levels[2,3], though this has never been confirmed experimentally. Here we exploit the design freedom afforded by metamaterials to fabricate active structures that exhibit the first experimental evidence of negative static electric susceptibility. Unlike the systems envisioned previously—which were expected to require reduced temperature and pressure[4]—negative values are readily achieved at room temperature and pressure. Further, values are readily tuneable throughout the negative range of stability $-1 < \chi^{(0)} < 0$, resulting in magnitudes that are over one thousand times greater than predicted previously[4]. This opens the door to new technological capabilities such as stable electrostatic levitation.**

Although static magnetic susceptibility may take positive or negative values in paramagnetic and diamagnetic materials respectively, a standard theoretical argument by Landau *et al.* (ref. 1, §14) suggests that static electric susceptibility must always be positive. More precisely: the real scalar describing the static electric susceptibility of an isotropic material must be positive, and, in general, the real symmetric second-rank tensor describing the static electric susceptibility must exhibit positive values for all three of its principal components. (An equivalent statement is that the presence of a material in the electric field of two conductors always increases, never decreases, their static mutual capacitance when compared to vacuum, regardless of the shape or orientation of the piece of material.) There is no reason to suppose that Landau *et al.*'s argument does not apply as rigorously to metamaterials as it does to any other type of material. Indeed, using an alternative theoretical argument, Wood and Pendry arrive at the same conclusion when considering metamaterials[5]. However, it must be borne in mind that, since such arguments implicitly assume that the material is in thermodynamic equilibrium, they do not necessarily hold for materials that are not in thermodynamic equilibrium—as noted, for example, in refs 2,3.

A general and apparently quite rigorous lower bound for the static electric susceptibility is provided by an elementary consideration of an electrical circuit consisting of a dc power supply, a resistor, and a capacitor in series: a hypothetical material with true negative static dielectric permittivity (static electric susceptibility less than minus one) would, when placed in the capacitor, lead to unphysical instabilities in the quasi-static limit (see, e.g., ref. 6 for a discussion in the context of dispersion in active and passive metamaterials). As a result, static electric susceptibility values less than minus one appear to be ruled out whether the material is in thermodynamic equilibrium or not. The question of negative static electric susceptibility values greater than minus one but less than zero is not addressed by such arguments, and it is negative values in this range that we consider herein. The possibility of negative static electric susceptibility in metastable systems with inverted populations has been discussed tentatively by Sanders[2,7] and predicted unequivocally by Chiao *et al.*[3,4,8,9]: the latter authors provide a quantitative estimate for the value of the static electric susceptibility of $\chi^{(0)} = -3.15 \times 10^{-4}$ at a pressure of 1 Torr and a temperature of 180 K in ammonia gas pumped by a carbon dioxide laser[4,8]. This prediction has never been experimentally verified.

We note that the real parts of the principal components of the complex electric susceptibility tensor may readily take negative values for periodic fields, associated with the phase difference between the electric field and the electric polarisation, and we emphasise that our interest herein is purely in the *static* case. We also note that we consider the linear electric susceptibility (polarisation proportional to electric field) as it pertains to a nonrelativistic, macroscopic, and homogeneous sample of material under the action of an electric field created by external charges. This may be considered the conventional interpretation of the electric susceptibility and is, for example, consistent with the meaning ascribed to the term by Landau *et al.*[1] and with the definition of the relative permittivity (via $\varepsilon = \chi + 1$) according to current ASTM standards[10]. There are a number of instances where negative static electric susceptibility or permittivity have been discussed in the literature in relation to quantities that do not correspond to this interpretation; for example, Kirzhnits *et al.* have shown that static permittivity may be negative in the sense that, if spatial dispersion is taken into account, the longitudinal permittivity at zero frequency but nonzero wavevector may exhibit negative values[11-14]. However, Kirzhnits *et al.*'s scenario concerns the situation where charge sources are placed within the material itself and, for the case of external test electrodes, Kirzhnits *et al.* reaffirm the conclusions of Landau *et al.* Herein we seek a 'true' negative static electric susceptibility in a conventional sense, but employ unconventional materials to achieve it, *viz.* active materials. It has previously been established that active metamaterials may exhibit electromagnetic *wave* behaviour not possible in their passive counterparts (e.g., refs 15-17); herein, the novelty is in applying the concept of active metamaterials to generate a new electro*static* material property.

The challenge is to create a material that polarises in essentially the opposite direction to normal under the action of a static electric field. For simplicity, we focus on creating an anisotropic material for which one principal component, $\chi_z^{(0)}$, of the effective static susceptibility tensor is negative, i.e., for which the induced polarisation is in the opposite direction to an electric field applied along the $\pm z$-axis. The general design concept is as follows: each unit cell, or 'meta-atom', consists of: (1) a mechanism to 'detect' the local electric field, (2) a system of conductors that may be charged to create an artificial dipole in a direction opposite to that of the electric field, (3) a mechanism by which the conductors may be charged in proportion to the detected electric field, and (4) a means by which to supply the energy necessary to do so. If the magnitude of the artificial dipole due to the charged conductors is sufficiently large with respect to the natural polarisation of the materials from which the meta-atom is made, the meta-atom will exhibit a net dipole moment in the opposite direction to normal and have a negative net electric polarisability. On a sufficiently large length-scale, many such meta-atoms may be considered a homogeneous medium with $\chi_z^{(0)}$ < 0. Given that all materials are inhomogeneous on an atomic length-scale, the description of the macroscopic behaviour of a macroscopic sample of such a metamaterial using the quantity $\chi_z^{(0)}$ is no less rigorous than for any other, 'conventional', material.

Here we report on two experimental implementations that provide evidence in support of this concept. In the first implementation, a metamaterial structure composed of a 15×15 array of meta-atoms with uniform and externally-controlled artificial polarisation was fabricated using etched copper-clad epoxy laminate boards, as shown in Fig. 1a&b (see also Methods and Supplementary Fig. 1 for a full specification). Each meta-atom contains two copper discs that may be charged by applying an artificial potential difference $V_p$ across them, creating an artificial dipole moment in the ±z direction. If we choose to apply $V_p$ in proportion to the potential difference $V_e$ across the external test electrodes such that $V_p = \alpha V_e$, we create artificial dipoles in the material whose response mimics natural linear electric polarisation, except with a readily-tuneable polarisability that is dependent on the value of $\alpha$.

$\chi_z^{(0)}$ was determined via static capacitance measurements on the external electrodes. Care was taken to employ a true dc method rather than simply an ac method at low frequencies (Methods). Results for various values of $\alpha$ are plotted in Fig. 1e, from which it is clear that $\chi_z^{(0)}$ decreases linearly with $\alpha$ and can attain negative values for sufficiently large $\alpha$: that is, for an inverted, artificial, polarisation of sufficiently large magnitude.

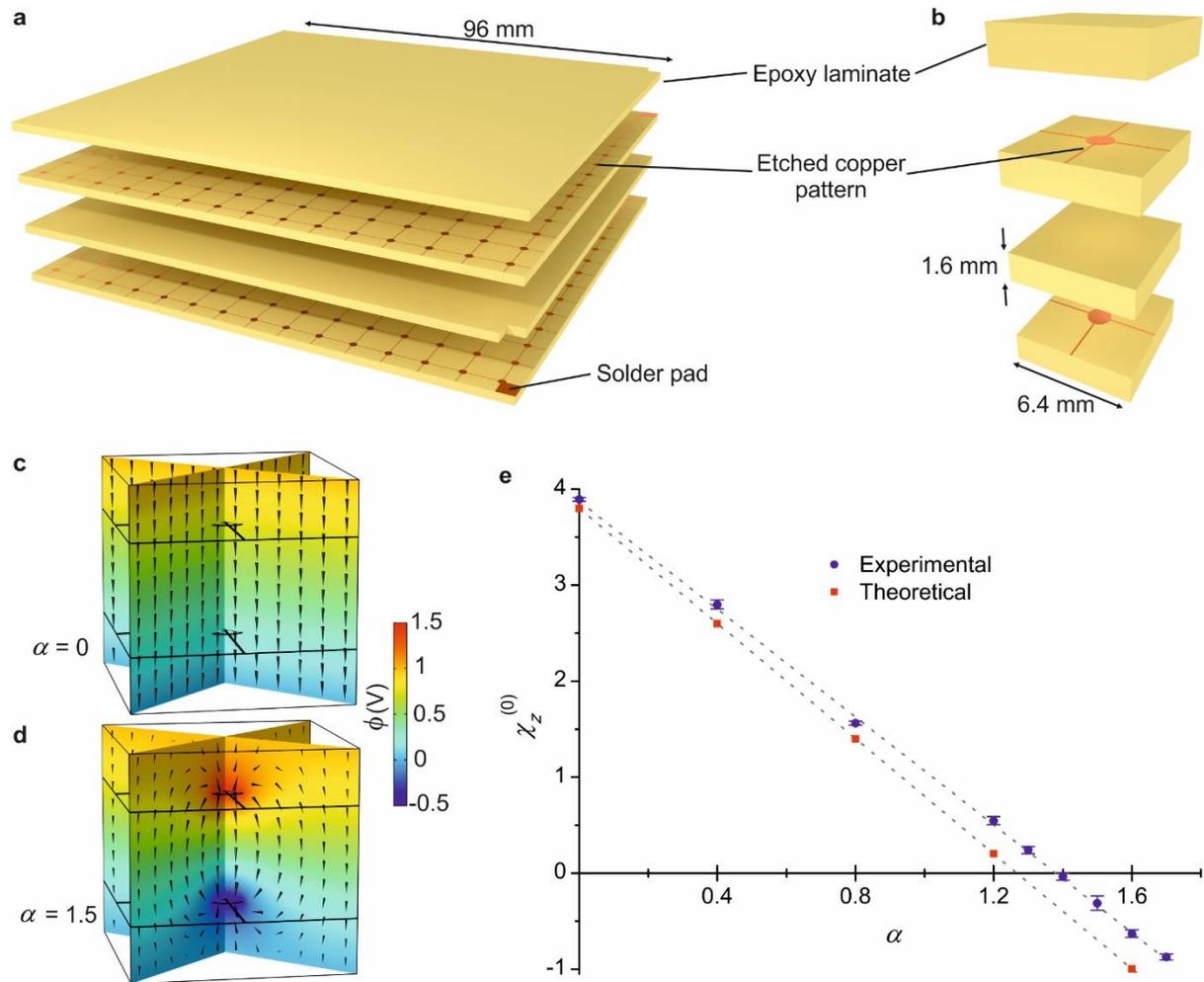

**Figure 1 | First metamaterial implementation.** Exploded view schematic diagrams of, **a**, the 15×15 metamaterial array formed using stacked epoxy laminate boards with etched copper, and, **b**, the structure of a single meta-atom. **c**, **d** Theoretical simulations showing the electric potential $\phi$ and the electric field (black cones) within a meta-atom for reduced driving voltages of $\alpha = 0$ and $\alpha = 1.5$ respectively. The latter exhibits an artificial dipole moment due to the charge on the copper discs and corresponds to a $\chi_z^{(0)} < 0$ state. **e** Experimental and theoretical data showing that $\chi_z^{(0)}$ decreases linearly with $\alpha$ and obtains negative values for sufficiently large $\alpha$. Experimental data points represent the mean of four repeated measurements and error bars represent one standard deviation uncertainties on the mean.

These experimental results are readily supported by theoretical modelling (Methods). For $\alpha = 0$, i.e., for no artificial dipole applied, the metamaterial behaves essentially as a homogeneous slab of epoxy laminate since the copper discs and tracks are at their 'natural' potentials and constitute a negligible volume of the meta-atom. In this case, the modelled electric field within the meta-atom may be seen to be essentially uniform, Fig. 1c, and the predicted value of $\chi_z^{(0)}$ is essentially that of the bulk FR4 epoxy laminate itself. For $\alpha > 0$, the artificial dipole is apparent in theoretical plots of the electric potential and electric field, Fig. 1d, and the theoretical value of $\chi_z^{(0)}$ is reduced, Fig. 1e. The origin of the quantitative discrepancy between the theoretical and experimental data apparent in Fig. 1e is currently unknown. However, it is clear that the theoretical model supports the qualitative features of the experimental results and, in particular, corroborates the existence of negative static permittivity for sufficiently large $\alpha$.

While the above implementation provides some evidence in support of the general concept, it cannot be considered a 'material' in any reasonable sense—not least, because the active voltage must be chosen and applied by hand. In a second implementation, we created a meta-atom that responds autonomously to the external electric field, Fig. 2 (see also Methods and Supplementary Figs. 2&3). It employs the simplest possible field-sensing element: two conductors, the 'sense electrodes', which, when subject to a vertical electric field produced by the external electrodes, are naturally raised to different electric potentials. The potential difference across the sense electrodes $V_s$ is detected by an instrumentation amplifier of gain $G$ which, in response, applies an amplified potential difference $G \times V_s$ across the drive electrodes.

For this second metamaterial, $\chi_z^{(0)}$ was again determined experimentally via dc capacitance measurements (Methods). With the metamaterial inserted between the external electrodes, the charge $Q$ on the upper external electrode increases linearly with the test voltage $V_e$, Fig. 2b, and the gradient provides a value of $C = (339.9 \pm 0.9)$ pF for the mutual capacitance of the external electrodes. With the metamaterial subsequently removed and the average separation of the external electrodes set to the same value, the $Q(V_e)$ data gives the empty capacitance $C_0 = (669.8 \pm 0.6)$ pF, Fig. 2b. Thus, it is seen that the mutual capacitance of the test electrodes is reduced by the presence of the metamaterial, and therefore the effective static electric susceptibility must be negative. A value for $\chi_z^{(0)}$ may be calculated most simply via $\chi_z^{(0)} \equiv C/C_0 - 1$, giving the result $\chi_z^{(0)} = -0.49$. The combined statistical and systematic uncertainty on this value is estimated to be 2% (Methods). Here, $G$ was deliberately chosen to produce a value of $\chi_z^{(0)}$ that is approximately in the middle of the region of interest $-1 < \chi_z^{(0)} < 0$. The magnitude of this value is of the order of $10^3$ times greater than that predicted for pumped ammonia gas[4]. By changing $G$, the value of $\chi_z^{(0)}$ could be readily tuned (Supplementary Fig. 4).

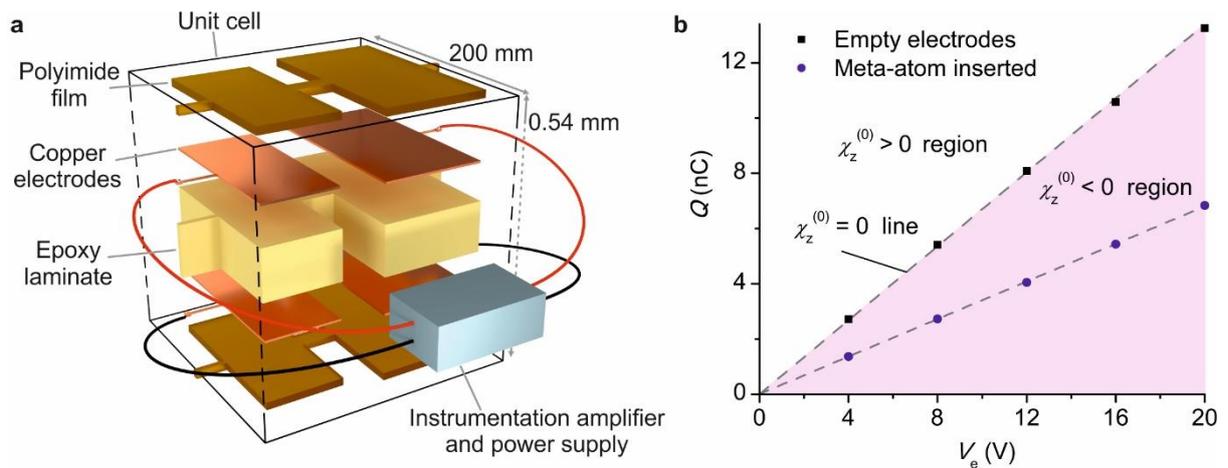

**Figure 2 | Second metamaterial implementation. a**, Exploded view schematic diagram of the meta-atom. The two copper electrodes on the left constitute the sense electrodes and the two on the right the drive electrodes. Relative vertical thicknesses of components within the unit cell are drawn ×100 for clarity. **b** Experimental data

for the charge $Q$ on the external test electrodes as a function of potential difference $V_e$ applied to them for the empty test electrodes (air as electric), and with the meta-atom inserted. Data points represent the mean of four repeated measurements. One standard deviation uncertainties on the mean values are in the range 0.01–0.04 nC, which is within the thickness of the data symbols.

The second metamaterial implementation again falls short of the general concept in the respect that the amplifier and the power supply are external to the unit cell. However, these components could, in principle, be implemented within the unit cell itself using, for example, integrated chip instrumentation amplifiers and coin cell batteries (though a somewhat thicker unit cell than that reported above may be required in practice). A further limitation is that the second, autonomous, metamaterial implementation concerns only a single meta-atom. In moving to multiple autonomous meta-atoms, one may expect that the meta-atom structure would require practical modification, but the response should remain stable for susceptibilities in the range $-1 < \chi_z^{(0)} < 0$. Similarly, with appropriate further engineering, materials with more than one negative principal component of $\chi^{(0)}$, and, in particular, isotropic materials with $\chi^{(0)} < 0$, should be possible.

Apart from the unique property of reducing the static mutual capacitance of two conductors, demonstrated above, materials with negative static electric susceptibility may be expected to exhibit some novel behaviour of potential technological interest. For example, negative static electric susceptibility is required for an electromagnetic cloak of the type originally considered by Pendry *et al.*[18] that operates for static electric fields. Alternatively, since negative static electric susceptibility materials are, in some sense, the electric analogues of diamagnetic materials, they may be expected to be capable of *stable* electro*static* levitation in analogy with well-known diamagnetic levitation effects; this may manifest, in principle, either as the levitation of a piece of negative static electric susceptibility material in a static electric field (as considered in refs 2,7,19 and analogous to the levitation of a diamagnetic material in a static magnetic field[20-22]), or as the levitation of a ferroelectric body above a piece of negative static electric susceptibility material (as considered in refs 3,4 and somewhat analogous to the levitation of a ferromagnet above a superconductor[23]). Further, a charged body may be expected to be capable of levitation above or inside a piece of negative static electric susceptibility material, with potential application in a new type of particle trap (as proposed in refs 3,4 and with no direct magnetic analogue due to the absence of magnetic monopoles). The realisation of active metamaterial structures reported herein, exhibiting evidence of negative static electric susceptibility with large magnitudes and at room temperature and pressure, brings such possibilities a step closer.

**Methods**

All measurements were carried out at room temperature and pressure.

**First metamaterial design.** Boards were fabricated from 1.6 mm FR4 dielectric with 18 μm copper cladding using standard photolithography techniques. Full design specifications are given in Supplementary Fig. 1. The design radius of the copper discs was $r \equiv a/10 = 0.64$ mm and the design width of the copper tracks was $w \equiv a/100 = 64$ µm, where $a = 6.4$ mm is the lateral dimension of the meta-atom. Given that the as-fabricated dimensions of small etched copper features may differ from the design dimensions, and to ensure the most accurate possible comparison between experiment and theory in this respect, the radius of the copper discs and the width of the copper tracks were measured experimentally using optical microscopy with a stage micrometer; the mean and standard uncertainty on the mean for 12 repeated measurements of each quantity, sampled randomly over the two patterned boards, were $r = (0.632 \pm 0.003)$ mm and $w = (44 \pm 3)$ µm respectively. The mean experimental values $r = 0.632$ mm and $w = 44$ µm were used as input to the theoretical model. Measurements using digital calipers and a micrometer screw gauge indicated no significant discrepancy from the nominal values for the FR4 board thickness, copper cladding thickness, or $a$, and nominal values were used as input for the theoretical model in these cases. In particular, the total thickness of the first metamaterial structure, including the outer external electrode layers, was found using electronic calipers to measure $(6.452 \pm 0.025)$ mm (eight repeated measurements), consistent with the nominal value of 6.472 mm (four layers of 1.6 mm FR4 and four layers of 18 µm copper); this implies that no significant gaps exist between the layers and justifies the absence of gaps in the theoretical model.

The discs in each layer were connected by thin copper tracks which allowed them to be raised conveniently to the same potential using external power supplies, but the tracks do not otherwise affect the qualitative behaviour of the device (as confirmed by theoretical modelling, Supplementary Information). An approximately uniform test electric field in the –$z$ direction was applied using external plane parallel test electrodes, implemented most simply by retaining the copper cladding on the outer surfaces of the uppermost and lowermost boards.

**Modelling of first metamaterial implementation.** Electrostatic modelling was carried out by solving Laplace's equation for the electric potential in a single meta-atom using the commercial software COMSOL (version 5.3, AC/DC Module, Electrostatics interface). The .mph file used to generate the theoretical data shown in Figs. 1c–e and Supplementary Fig. 5 is included as Supplementary Model 1. The epoxy laminate boards were treated as homogeneous and isotropic dielectric layers with a known intrinsic static electric susceptibility (= 3.9), and the copper regions were treated as perfectly-conducting domains on whose boundaries the potential was specified according to the externally applied voltages. The cross-sections of the copper features were assumed to be rectangular, i.e., no attempt was made to account for etching undercut. With $V_e$ set to the arbitrary value of 1 V throughout, the charge $Q_1$ on the top and bottom surfaces of the meta-atom for a given value of $\alpha$ was determined by the inbuilt COMSOL 'Terminal charge' function. From this, the mutual capacitance per meta-atom of the upper and lower external electrodes was determined according to $C_1 = Q_1/V_e$. Knowing $C_1$, the effective static electric susceptibility may be calculated from $\chi_z^{(0)} = C_1 d/(\varepsilon_0 a^2) - 1$, where $\varepsilon_0$ is the permittivity of free space and $d = 6.436$ mm is the thickness of the metamaterial (four 1.6 mm FR4 boards and two internal layers of 18 μm copper for the discs and tracks).

With the boundary conditions on the lateral surfaces of the meta-atom set to 'Zero charge', i.e., with no lateral component of the electric induction field **D**, the model considers the situation where a given meta-atom is surrounded by identical meta-atoms. We believe this constitutes the best approach to determining the 'true model value' of $\chi_z^{(0)}$. In particular, the value of $\chi_z^{(0)}$ thus extracted may be expected to be independent of the number of meta-atoms considered and it is appropriate to consider a single meta-atom for computational simplicity. This was confirmed explicitly by generating equivalent models for two and four meta-atoms (Supplementary Models 2 and 3 respectively); for the given mesh settings, these models return the same value of $\chi_z^{(0)}$ as the single meta-atom model to within 0.1 %

The computational accuracy of the single meta-atom model was validated by investigating the convergence of the values of $\chi_z^{(0)}$ with regard to the size of the finite element mesh (Supplementary Information) and the model data plotted in Fig. 1e represent the extrapolated, converged values. Convergence plots are reported in Supplementary Fig. 5.

**Experimental determination of $\chi_z^{(0)}$ for the first metamaterial implementation.** A schematic diagram of the experimental setup for the first metamaterial implementation is shown in Supplementary Fig. 6. With the lower external electrode grounded and the upper external electrode raised to potential $V_e$, the lower and upper discs would, if floating, be raised to the potentials $V_e/4$ and $3V_e/4$ respectively. To create the artificial dipole moment, we add to this an additional potential difference $V_p$ across the discs by applying $V_l = V_e/4 - V_p/2$ to the lower discs and $V_u = 3V_e/4 + V_p/2$ to the upper discs. Three separate high voltage dc power supplies were used to apply $V_l$, $V_u$, and $V_e$. With the bottom external electrode grounded throughout, the power supplies to the lower and upper discs were switched on, applying potentials $V_l$ and $V_u$ respectively, before the upper external electrode was raised to potential $V_e$ by temporarily touching the output lead from the third dc power supply to the upper copper surface, depositing on it charge $Q$. After switching off the power supplies to the lower and upper discs, $Q$ was determined by temporarily touching the lead from the electrometer to the upper external electrode. The mutual capacitance of the external electrodes was then calculated via $C \equiv Q/V_e$. Throughout the experiment, $V_e$ was kept fixed at 1 kV and $\alpha$ was varied by modifying $V_l$ and $V_u$ accordingly.

To extract a value for $\chi_z^{(0)}$ it is necessary to use the lateral dimension of the metamaterial structure. Again, given the as-fabricated dimension may differ from the design dimension of $L \equiv 15a = 96$ mm, and to ensure the most relevant application of the model, this length was measured experimentally; using digital calipers, the mean and standard uncertainty for eight repeated measurements were $L = (95.86 \pm 0.02)$ mm, and the mean experimental value of $L = 95.86$ mm was used as input to the model.

An approximate value for $\chi_z^{(0)}$ could, in principle, be extracted from the experimental values of $C$ via the elementary formula for plane parallel electrodes $\chi_z^{(0)} = Cd/(\varepsilon_0 L^2)-1$. However, this approach would ignore fringing fields, which, for the particular electrode geometry employed in this first metamaterial implementation, are not insubstantial: for an air-filled capacitor with $d = 6.436$ mm and $L = 95.86$ mm, the total capacitance, including fringing fields, is 22 % greater than that predicted by the formula $C = \varepsilon_0 L^2/d$ (Supplementary Model 4). Therefore, a more accurate approach, which accounts for fringing fields, was implemented instead using a finite element simulation to facilitate the extraction as follows. First, Supplementary Model 4 was used to calculate the values of $C$ for a homogeneous dielectric at a range of susceptibilities: $\chi = -0.9, -0.5, 0, 1, 2, 3,$ and $4$. On a plot of $\chi(C)$, Supplementary

Fig. 7, these data points indicate a linear relationship between $\chi$ and $C$ that may be fitted by the straight line $\chi = \beta C + \gamma$, where $\beta = 7.896 \times 10^{10}$ F$^{-1}$, and $\gamma = -1.210$. Thus, the value of $\chi_z^{(0)}$ for a given value of $C$ may be extracted using the empirical formula $\chi_z^{(0)} = 7.896 \times 10^{10} \hat{C} - 1.210$, where $\hat{C}$ is the numerical value of the measured capacitance in Farads. The computational accuracy of Supplementary Model 4 was validated by comparing the predicted capacitance in the $\chi = 0$ case (no material between plates) with a known literature value (Supplementary Information).

This procedure ignores the effects on the measured capacitance of the Kapton tape holding the boards together, the solder pads and associated cutaway squares, the attached wires, and stray capacitances to ground.

**Second metamaterial design.** The main structure was fabricated from 0.4 mm FR4 dielectric with 18 μm copper cladding using standard photolithography techniques and laminated using 50 μm polyimide film; full design specifications are given in Supplementary Fig. 2. The size of the unit cell is defined by the cuboidal volume between the external electrodes, which were square with edge length $L = 200$ mm (negligible uncertainty) and, for all quoted measurements, at a separation of $d = (0.54 \pm 0.01)$ mm. A relatively large lateral dimension and small separation were employed for the following practical reasons: (1) large area external electrodes at narrow separation increased the measured capacitance such that test voltages ~ 1–10 V created charges ~ 1–10 nC on the external electrodes, which could be readily measured using a basic electrometer setup (unlike for the first metamaterial implementation which could exploit high voltages ~ 1kV to create large charges, applied voltages in the second implementation were limited by the 36 V maximum operating voltage of the particular operational amplifiers employed), (2) the area of the sense electrodes was required to be sufficiently large that, upon application of the external electric field, the potential difference across them would not be quickly neutralised due to a small amount of charge flow caused by the non-ideal operation of the instrumentation amplifier, and (3) large area external electrodes at narrow separation reduced the effects of fringing fields (a standard empirical formula—equation (50) of ref. 24—indicates that for $L = 200$ mm and $d = 0.54$ mm, fringing fields account for 1% of the total capacitance when air-filled and 2% when filled with a $\chi_z^{(0)} = -0.5$ material).

The instrumentation amplifier was based on a standard design (e.g., ref. 25) and the exact circuit diagram is shown in Supplementary Fig. 3b.

**Experimental determination of $\chi_z^{(0)}$ for the second metamaterial implementation.** A schematic diagram of the experimental setup for the second metamaterial implementation is shown in Supplementary Fig. 3a. With the positive output lead from the dc power supply to the external electrodes initially touching the wire connected to the upper external electrode, the dc power supply was switched on, applying potentials $+V_e/2$ and $-V_e/2$ to the upper and lower external electrodes respectively. The positive output lead was then removed from touching the wire connected to the upper external electrode, leaving charge $Q$ on the upper external electrode. After switching off the power supply, $Q$ was determined by temporarily touching the lead from the electrometer to the upper external electrode. Four grounding switches $GS_1$, $GS_2$, $GS_3$, and $GS_4$, were used to ensure that the sense and drive electrodes were at ground potential prior to the measurements and were opened just before each measurement. The gain $G$ of the instrumentation amplifier was tuned by changing the resistance $R_g$ according to the standard expression $G = 1+2R/R_g$, where $R$ and $R_g$ are defined with respect to Supplementary Fig. 3.

Unlike for the first metamaterial, external test electrodes were employed that could be set to a variable separation using micrometer screw brackets, enabling the mutual capacitance of the external electrodes to be experimentally determined with no material placed between them (air as dielectric). The separation of the external electrodes is crucial to the extracted value of $\chi_z^{(0)}$ and the following method ensured that plate separations of $d \approx 0.5$ mm could be determined to within $\approx 2$ % uncertainty. A copper transmission electron microscopy support grid was used as a small 'ruler' by attaching it to the edge of the external electrodes using a piece of clear sticky tape and photographing it using a mobile phone camera with an attached macro lens. Analysing the image, the separation of the external electrodes could be determined using the known support grid feature dimensions for calibration. Quoted values of $d$ below refer to the mean, and standard deviation on the mean, of six separate measurements: two different support grids employed at three different locations on the perimeter of the external electrodes. External electrode structures with sufficient rigidity and flatness were made by adhering 0.8 mm FR4 board with single-sided copper cladding to $200 \times 200 \times 80$ mm$^3$ blocks of machined tooling foam.

With the metamaterial inserted such that it rested on the lower external electrode and the upper external electrode rested on the metamaterial, the plate separation was measured as $d = (0.54 \pm 0.01)$ mm and the $Q(V_e)$ data shown in Fig. 2b was recorded. The measured separation matches, to within experimental error, the nominal structure thickness of 0.536 mm (one layer of 0.4 mm FR4, two layers of 18 µm copper, and two layers of 50 µm polyimide film), indicating that the technique for the measurement of $d$ is free from significant systematic uncertainties at the quoted 2% level of random uncertainty. With the metamaterial subsequently removed, the plate separation was set, measured, and adjusted until an average separation of $d = (0.54 \pm 0.01)$ mm was again recorded, and the $Q(V_e)$ data for the empty electrodes shown in Fig. 2b was taken.

The systematic uncertainty in the determination of the static electric susceptibility may be investigated by using the empty-electrode results $C_0 = 669.8$ pF and $d = 0.54$ mm to 'measure' the permittivity of free space (here, and throughout, we ignore the $\approx 0.05\%$ difference between the permittivity of air and that of free space[26], which is essentially negligible for our purposes). Using a standard empirical formula—equation (50) of ref. 24—to take into account the 1% fringing field mentioned above gives a 'measurement' of $\varepsilon_0 = 8.93$ pF/m. This is within 1% of the defined value $\varepsilon_0 = 8.85\ldots$ pF/m, implying that the overall uncertainty in the measurement technique is probably dominated by the 2% statistical uncertainty in the measurement of $d$.

$\chi_z^{(0)}$ for the metamaterial may be determined independently of the empty-electrode results by using the measurements $C = (339.9 \pm 0.9)$ pF and $d = (0.54 \pm 0.01)$ mm. Employing the standard empirical formula of ref. 24 to take into account the 2% fringing field, and using standard propagation of statistical uncertainties (dominated by the uncertainty on $d$), gives the result $\chi_z^{(0)} = -0.49 \pm 0.01$. This justifies the more direct method used in the main text (which ignores fringing fields) and produces the stated 2% uncertainty.

**Acknowledgements**

This work was funded by the Engineering and Physical Sciences Research Council UK (Grant No. EP/I034548/1). The authors thank G. Cook, P. Flaxman, P. Pattinson, G. Ring and R. Vincent for technical support, and E. Liotti, J. B. Pendry, P. Shrimpton and S. C. Speller for useful discussions.

**Author Contributions**

F.C. conceived the idea, performed the experiments, and carried out the theoretical modelling. J.A.J.F. and D.I. assisted in the design of the circuitry for the second metamaterial. A.W. and S.M.M provided equipment. F.C. wrote the paper with input from all the authors. F. C. and P.S.G. co-directed the research.

**Correspondence and requests for materials** should be addressed to F.C. or P.S.G.